\documentclass[11pt]{article}
\usepackage{graphicx,caption,subfigure}
\makeatletter
\newcommand{\singlespacing}{\let\CS=\@currsize\renewcommand{\baselinestretch}{1.0}\tiny\CS}
\newcommand{\doublespacing}{\let\CS=\@currsize\renewcommand{\baselinestretch}{1.5}\tiny\CS}

\begin{document}
\title{Pion, Kaon and Antiproton Production in $Pb+Pb$ Collisions at LHC Energy $\sqrt{s_{NN}}$ = 2.76 TeV : A Model-based Analysis}
\author{P.
Guptaroy$^1$\thanks{e-mail: gpradeepta@gmail.com (Communicating author)}, S. Guptaroy$^2$\thanks{e-mail: simaguptaroy@yahoo.com}\\
{\small $^1$ Department of Physics, Raghunathpur College,}\\
 {\small P.O.: Raghunathpur 723133,  Dist.: Purulia (WB), India.}\\
 {\small $^2$ Department of Physics, Basantidevi College,}\\
 {\small 147B Rashbehari Avenue, Kolkata 700029 India.}}
\date{}
\maketitle
\begin{abstract}
Large Hadron Collider (LHC) had produced a vast amount of high precision data for high energy heavy ion collision. We attempt here to study (i) transverse momenta spectra, (ii)  $K/\pi$, $p/\pi$ ratio behaviours, (iii)rapidity distribution, and (iv) the nuclear modification factors of the pion, kaon and antiproton produced in $p+p$ and $Pb+Pb$ collisions at energy $\sqrt{s_{NN}}$ = 2.76 TeV, on the basis of Sequential Chain Model (SCM). Comparisons of the
model-based results with the measured data on these observables are generally found to be modestly satisfactory.
\end{abstract}

\newpage
\section{Introduction}
Heavy ion collisions at ultra relativistic energies might produce a new form of QCD matter characterized by the
deconfined state of quarks and gluons (partons) \cite{alice1401}. Measurements of the production of identified particles
provide information about the dynamics of this dense matter. The yield of identified hadrons, their
multiplicity distributions, as well as the rapidity and transverse momentum spectra are the
basic observables in proton-proton and heavy ion collisions at any energy regime, from a few
GeV per nucleon to the new ultra-relativistic LHC regime, spanning c.m. energies of a few TeV.\cite{riggi} Recently, experimental results in
$Pb+Pb$ collisions at $\sqrt {s_{NN}}$ = 2.76 TeV in the Large Hadron Collider (LHC)
are also reported by the different groups. These results might provide another opportunity to investigate the bulk properties
of exotic QCD matter, the so-called QGP-hypothesis. But the exact nature of QGP-
hadron phase transition is still plagued by uncertainties.\cite{zhang}
\par
Our objective in this work is to use a model, known as `Sequential Chain Model (SCM)', which is different from `standard' framework, in interpreting the transverse momenta spectra, some ratio-behaviours, rapidity spectra and the nuclear modification factor of the pions, kaons and antiprotons for $p+p$ and $Pb+Pb$ collisions at LHC energy $\sqrt{s_{NN}}$ = 2.76 TeV. Very recently, a question has been raised about the quark-gluon composition of proton.\cite{deflo} So, in order to explain the huge amount of heavy ion collision data, we put forward this model which has no QGP-tag and is different from the quark-hypothesis.
\par
The organization of this work is as follows. In section 2
we give a brief outline of our approach, the SCM. In the next section (section 3) the results arrived
at have been presented with table and figures. And in
the last section (section 4) we offer the final remarks and
conclusions.
\section{Outline of the Model}
This section is divided by two subsections (1) the basic model in $p+p$ collision and (2) subsequent transformation to the $A+B$ collisions.
\subsection{The Basic Model: An Outline}
According to
this Sequential Chain Model (SCM), high energy hadronic
interactions boil down, essentially, to the pion-pion interactions;
as the protons are conceived in this model as
$p$ ~ = ~ ($\pi^+$$\pi^0$$\vartheta$), where $\vartheta$ is a spectator particle needed for
the dynamical generation of quantum numbers of the nucleons.\cite{pgr14}-\cite{bhat882}
\par
 The inclusive cross-section of the $\pi^-$-meson produced in the
$p+p$ collisions at high energies has been calculated by using field theoretical calculations and Feynman diagram techniques with the infinite momentum frame approximation method. The inclusive cross-section is given by the undernoted relation \cite{pgr14}-\cite{bhat882}
\begin{equation}\displaystyle E\frac{d^3\sigma}{dp^3}|_{pp \rightarrow
\pi^- x}  \cong \Gamma_{\pi^-} \exp(- 2.38 <n_{\pi^-}>_{pp}
x)\frac{1}{p_T^{(N_R^{\pi^-})}} \exp(\frac{-2.68
p_T^2}{<n_{\pi^-}>_{pp}(1-x)})   ~ ,
\end{equation}
with \begin{equation}\displaystyle {<n_{\pi^+}>_{pp} ~ \cong  ~
<n_{\pi^-}>_{pp} ~ \cong
 ~ <n_{\pi^0}>_{pp}  ~ \cong  ~ 1.1s^{1/5} ~,}
 \end{equation}
 where $\Gamma_{\pi^-}$ is the
normalisation factor which will increase as the inelastic
cross-section increases and it is different for different energy
region and for various collisions. The terms
$p_T$, $x$ [$x ~ \simeq  ~ 2p_{z{cm}}/{\sqrt s} ~ = ~ 2m_T \sinh{y_{cm}}/{\sqrt s}$] in equation (1) represent the transverse momentum,
Feynman Scaling variable respectively. The $s$ in equation (2) is the square of the c.m. energy.
\par
$1/p_T^{N_R^{\pi^-}}$ of the expression (1) is the `constituent
rearrangement term'. It arises out of the partons inside the proton. At high energy interaction processes the partons undergo some dissipation losses due to impact and impulses of the projectile on the target.
This term essentially provides a damping term in terms of a power law. The exponent of $p_T$, i.e. $N_R$, varies on both the
collision process and the specific $p_T$-range. We have to parametrize this term with the view of two
physical points, viz., the amount of momentum transfer and the
contributions from a phase factor arising out of the rearrangement
of the constituent partons. The relation for $N_R$ is to be given by \cite{pgr08}
\begin{equation}\displaystyle
N_R=4<N_{part}>^{1/3}\theta,
\end{equation}
where $<N_{part}>$ denotes the average number of participating
nucleons and $\theta$ values are to be obtained phenomenologically
from the fits to the data-points.
\par
Similarly, for kaons of any specific variety ( $K^+$, $K^-$, $K^0$
or $\bar{K^0}$ ) we have
\begin{equation}
\displaystyle E \frac{d^3\sigma}{dp^3}|_{pp \rightarrow K^- x}
 ~ \cong  ~ \Gamma_{K^-}\exp(  -  6.55 <n_{K^-}>_{pp}
 x) ~ \frac{1}{p_T^{(N_R^{K^-})}}\exp(\frac{-  1.33
  p_T^2}{<n_{K^-}>^{3/2}_{pp}}) ~ ~ ,
\end{equation}
 with
\begin{equation}
\displaystyle <n_{K^+}>_{pp}   \cong  <n_{K^-}>_{pp}  \cong
  <n_{K^0}>_{pp}  \cong  <n_{\bar{K^0}}>_{pp} \cong
5\times10^{-2}  s^{1/4}  .
\end{equation}
And for the antiproton production in $pp$ collision at high
energies, the derived expression for inclusive cross-section is
\begin{equation}
\displaystyle E\frac{d^3\sigma}{dp^3}|_{pp\rightarrow{\bar p}x}
 ~ \cong \Gamma_{\bar p} \exp(-25.4 <n_{\bar{p}}>_{pp} x)\frac{1}{p_T^{({N_R}^{\bar p})}}\exp(\frac{-0.66 ((p_T^2)_{\bar
p}+ {m_{\bar p}}^2)}{<n_{\bar p}>^{3/2}_{pp} (1-x)})
 ~  ,
\end{equation}
with
\begin{equation}
\displaystyle{ <n_{\bar p}>_{pp}  ~ \cong <n_p>_{pp}  ~ \cong
 ~ 2\times10^{-2} ~ s^{1/4} ~ ,}
 \end{equation}
\subsection{The Path from $pp$ to $AB$ Collisions}
In order to transform the inclusive cross-section from $pp\rightarrow C^-+ X$ to $AB\rightarrow C^-+ X$ collisions ( here, $C^-$ stands for $\pi^-$, $K^-$ and $\bar p$, as the case may be), we use the undernoted relation; \cite{wong}
\begin{equation}\displaystyle
E\frac{d^3\sigma}{dp^3}|_{AB \rightarrow C^-+ X} \cong
{\frac{(A \sigma_B + B
\sigma_A)}{\sigma_{AB}}}{\frac{1}{1+a'(A^{1/3}+B^{1/3})}}{E\frac{d^3\sigma}{dp^3}}
|_{pp \rightarrow C^-+ X},
\end{equation}
Here, in the above equation [eqn.(8)], the first factor  gives
a measure of the number of wounded nucleons i.e. of the probable
number of participants, wherein $A\sigma_B$ gives the probability
cross-section of collision with `$B$' nucleus (target), had all the
nucleons of $A$ suffered collisions with $B$-target. And $B\sigma_A$
has just the same physical meaning, with $A$ and $B$ replaced.
Furthermore, $\sigma_A$ is the nucleon(proton)-nucleus(A)
interaction cross-section, $\sigma_B$ is the inelastic
nucleon(proton)-nucleus(B) reaction cross-section and $\sigma_{AB}$
is the inelastic $AB$ cross-section for the collision of nucleus $A$
and nucleus $B$. The values of $\sigma_{AB}$, $\sigma_{A}$,
$\sigma_{B}$ have been worked here out by
the following formula \cite{na5002}
\begin{equation}
\displaystyle{ \sigma^{inel}_{AB} ~ = ~ \sigma_{0} ~
(A^{1/3}_{projectile} + A^{1/3}_{target} - \delta)^2}
\end{equation}
with $\sigma_{0} = 68.8$ mb, $\delta= 1.32$.
\par
 The second term  in expression (8) is a physical factor
related with energy degradation of the secondaries due to multiple
collision effects. The parameter $a'$ occurring in this term is a
measure of the fraction of the nucleons that suffer energy loss. The
maximum value of $a'$ is unity, while all the nucleons suffer energy
loss. This $a'$ parameter is usually to be chosen \cite{wong},
depending on the centrality of the collisions and the nature of the
secondaries.
\section{The Results}
This section will be divided in the following sub-sections: (i) the $p_T$-spectra of pion, kaon and antiproton in both $p+p$ and $Pb+Pb$ collisions at $\sqrt{s_{NN}}$ = 2.76 TeV; (ii) $K/\pi$ and $p/\pi$ ratio behaviour at $Pb+Pb$ collisions at $\sqrt{s_{NN}}$ = 2.76 TeV; (iii) rapidity distribution of pion for the most central collisions of $Pb+Pb$ in the above-mentioned energy and (iv) the nuclear modification factor $R_{AA}$ in the same energy range.
\subsection{Transverse Momenta Spectra of Charged Hadron in $p+p$ and $Pb+Pb$ Collision at $\sqrt{s_{NN}}$ =2.76 TeV}
 We can write from expression (8), the transformed SCM-based transverse-momentum distributions
for $A+B\rightarrow C^-+X$-type reactions in the
following generalized notation:
\begin{equation}\displaystyle{
\frac{1}{2\pi p_T}}{\frac{d^2N}{dp_T dy}|_{A+B\rightarrow C^-+
x}=\alpha_{C^-}\frac{1}{p_T^{N_R^{C^-}}}\exp(-\beta_{C^-} \times
p_T^2).}
\end{equation}
Where, for example, the
parameter $\alpha_{\pi^-}$ can be written in the following form:
\begin{equation}\displaystyle{
\alpha_{\pi^-}={\frac{(A \sigma_B + B
\sigma_A)}{\sigma_{AB}}} {\frac{1}{1+a(A^{1/3}+B^{1/3})}}\Gamma_{\pi^-}\exp(- 2.38
<n_{\pi^-}>_{pp} x)}
\end{equation}
\par
In a similar way, the values of $\beta_{\pi^-}$ of the equation (10) have been calculated with the help of eqn.(1), eqn.(2). The values of $\alpha_{K^-}$, $\alpha_{\bar p}$, $\beta_{\pi^-}$ and $\beta_{\bar p}$ have been calculated accordingly by using eqns. (4)- (8). Moreover, for calculation for transverse momenta distribution of antiproton production, the the exponential part will be $\exp(-\beta_{\bar p}(p_T^2 + {m_{\bar p}}^2))$ ($m_{\bar p} ~ \sim ~ 938 ~ MeV/c^2$).
\par
$N_R^{C^-}$ of the expression (10) have been calculated by using eqn. (3).

\subsubsection{Production of $\pi^-$, $K^-$ and $\bar p$ in $p+p$ Collision at $\sqrt{s_{NN}}$ =2.76 TeV}
In table 1, the calculated values of $\alpha$, $N_R$ and $\beta$ of eqn.(10) for $\pi^-$, $K^-$ and $\bar p$ have been given.
\par
In figure (1), we have drawn the invariant yields against $p_T$ for $\pi^-$, $K^-$ and $\bar p$.
By using equation (10) and Table 1 we have plotted the solid lines against the
experimental background \cite{cms12}. The dotted lines in the Fig. show PYTHIA-based results \cite{cms12}.
\subsubsection{Invariant Yields of $\pi^-$, $K^-$ and $\bar p$ in $Pb+Pb$-Collision at $\sqrt{s_{NN}}$ =2.76 TeV}
In a similar fashion, the invariant yields of $\pi^-$, $K^-$ and $\bar p$ in $Pb+Pb$ collision at LHC energy $\sqrt s_{NN}$ =2.76 TeV for different centralities have been plotted in Fig.2(a), 2(b) and 2(c) respectively. The solid lines in the Fig. are the theoretical SCM results while the points show the experimental values.\cite{ALICE13}. The values of $\alpha_{C^-}$, $N_R^{C^-}$ and $\beta_{C^-}$ of eqn.(10) for pion, kaon and antiproton and for different centralities have been given in Table 2.
\subsection{The $K/\pi$ and $p/\pi$ ratios}
The  model-based $K/\pi$ and $p/\pi$ ratios as a function of $p_T$ at energies $\sqrt{s_{NN}}$ = 2.76 TeV
have been obtained from the expression (10), Table 1 and Table 2. Data in
Figs. 3(a) and 3(b), for different centralities, viz., for 0-5$\%$, 20-30$\%$ and 70-80$\%$, are taken from Ref. \cite{ALICE13}. Lines in the Figure show the theoretical plots.
\subsection{The Rapidity Distribution}
For the calculation of rapidity distribution, we can make use of the following standard relation \cite{abk},
\begin{equation}\displaystyle{
\frac{dN}{dy}=\frac{1}{\sigma_{in}}\int[E\frac{d^3\sigma}{dp^3}]d^2p_T}
\end{equation}
By using eqn. (1), eqn. (8), Table 2 and eqn. (12), we arrive at the SCM-based rapidity distribution, which is given hereunder;
\begin{equation}\displaystyle{
\frac{dN}{dy}=1895\exp(-0.025\sinh y_{cm})}.
\end{equation}
The $y_{cm}$ of the above eqn. (eqn.(13)) has come from $\exp(-2.38<n_{\pi^-}>x)$ of eqn. (11) with $x ~ = ~ 2m_T\sinh{y_{cm}}/\sqrt s$.
\par
In Fig.4, we have plotted theoretical $dN/dy$ versus $y$ with the help of above equation [eqn. (13)] against experimental background
\cite{ALICE132}. The dotted line in this Fig. shows the Gaussian fit \cite{ALICE132}.
\subsection{The Nuclear Modification Factor}
The nuclear modification factor (NMF) $R_{AA}$ is defined as ratio of charged particle yield in $Pb+Pb$ to that in $p+p$, scaled by the number of binary nuclear collisions $<N_{coll}>$ \cite{jacek} and is given hereunder
\begin{equation}\displaystyle{
R_{AA}(p_T)=\frac{(1/N^{AA}_{evt})d^2N^{AA}_{ch}/dp_T d\eta}{<N_{coll}>(1/N^{pp}_{evt})d^2N^{pp}_{ch}/dp_T d\eta},}
\end{equation}
where $<N_{coll}>$ is related with the average nuclei thickness function($<T_{AA}>$) by the following relation \cite{jacek}
\begin{equation}\displaystyle{
<N_{coll}>=<T_{AA}> \sigma_{pp}^{incl}.}
\end{equation}
Here, $\sigma_{pp}^{incl}$ is the total inelastic cross section of $p + p$ interactions.
\par
The $d^2N/dp_Td\eta$ is related to $d^2N/dp_Tdy$ by the following relation:\cite{wong}
\begin{equation}\displaystyle{
\frac{d^2N}{dp_Td\eta} ~ = ~ \sqrt{1-\frac{m^2}{m_T^2\cosh^2 y}}\frac{d^2N}{dp_Tdy}}.
\end{equation}
In the region, $y>>0$, $\frac{d^2N}{dp_Td\eta} \sim \frac{d^2N}{dp_Tdy}$.
\par
The SCM-based results on NMF's for $\pi^-$, $K^-$ and $\bar p$ in central $Pb+Pb$ collisions at
energies $\sqrt{s_{NN}}$ = 2.76 TeV
are deduced on the basis of Eqn.(10), Table 1 and Table 2. The equations in connection with $(R_{AA})_{\pi^-}$, $(R_{AA})_{K^-}$ and $(R_{AA})_{\bar p}$ are give by the following relations
and they are plotted
in Fig.5 against
$p_T$.
The solid lines in the figure show the theoretical results, while the points show the
experimentally  measured results \cite{marek};
\begin{equation}\displaystyle{
(R_{AA})_{\pi^-}=0.35 p_T^{0.33},}
\end{equation}
\begin{equation}\displaystyle{
(R_{AA})_{K}=0.40p_T^{0.73},}
\end{equation}
\begin{equation}\displaystyle{
(R_{AA})_{\bar p}=0.25p_T^{1.33}.}
\end{equation}
\section{Discussions and Conclusions}
Let us now make some comments on the results arrived at and shown by the diagrams on the case-to-case basis.
\par
(1) The invariant yields against transverse momenta ($p_T$) for $\pi^-$, $K^-$ and $\bar p$ in proton-proton collisions obtained on the basis of the SCM are shown in Figure 1. Except for very low-$p_T$ region, there is a bit degree of success. The model disagrees in the low-$p_T$ region. This is due to the fact that the model has turned essentially into a mixed one with the inclusion of power law due to the inclusion of partonic rearrangement factor. However, the power-law part of the equation might not be the only factor for this type of discrepancy. The initial condition and dynamical evolution in heavy-ion collisions are more complicated than we expect. Till now, we do not know the exact nature of reaction mechanism. One might take into account some other factors like radial flow or thermal equilibrium.
\par
(2) Similarly, in calculating the yields for different transverse momenta and for different centralities for $\pi^-$, $K^-$ and $\bar p$ in lead-lead collisions, we use eqns. (8), (9) and (10) alongwith eqns. (1)-(7). The results are given in Table 2 and are depicted in Figs. 2(a), 2(b) and 2(c) respectively. The top-most curves are for central collisions (0-5$\%$) and the lowest curves are peripheral ones (80-90$\%$). In between these two curves, other centralities (5-10$\%$, 10-20$\%$, 20-30$\%$, 30-40$\%$, 40-50$\%$, 50-60$\%$, 60-70$\%$ and 70-80$\%$) have been plotted. For the production of pions, the SCM-based results show good fits from central to peripheral collision. Slight disagreements observed at very low-$p_T$ regions for kaons and protons at central collisions. These are due to the power law part of the model. Here, we see that the constituent rearrangement terms are clearly centrality dependent. This explanation is also true for low-$p_T$ region data in $p+p$ collision.
\par
(3) The $K/\pi$ and $p/\pi$ ratio behaviours  for different centralities are calculated in the light if the SCM and they are plotted in Figs. 3(a) and 3(b) respectively. The theoretical $K/\pi$ ratio behaviours are in good agreement with experimental values. Some disagreement are observed in central $p/\pi$- ratio in low-$p_T$ regions.
\par
(4) In explaining the rapidity distribution for production of pions (Figure 4), the majority of the produced secondaries, the model works agreeably with data. The comparison with Gaussian fit is satisfactory.
\par
(5) The nuclear modification factors for pion, kaon and proton for central $Pb+Pb$-collisions for different transverse momenta have been calculated and they are plotted in Figure 5. While the theoretical plots are agreeable in low-$p_T$ regions, they disagree in high-$p_T$.
\par
Now, let us sum up our observations in the following points; (1) the model under consideration here explains the data modesly well on $Pb+Pb$ collisions at $\sqrt{s_{NN}}$ = 2.76 TeV. (2) The particle production in heavy ion collisions can be viewed alternatively by this Sequential Chain Model.
 \par
 {\bf{Acknowledgements}}\\
 The work is supported by University Grants Commission, India, against the order no. PSW-30/12(ERO) dt.05 Feb-13.

\newpage
\begin{table}
\caption{Values of $\alpha$, $N_R$ and $\beta$ for
pions, kaons, antiproton and proton productions in $p+p$ collisions at
$\sqrt{s_{NN}}$= 2.76 TeV}
\begin{center}
\begin{tabular}{ccc}
\hline
\hline
$(\alpha_{\pi^-})_{pp}$&$(N_R^{\pi^-})_{pp}$&$(\beta_{\pi^-})_{pp}$\\
0.581&2.163&0.30\\
\hline
$(\alpha_{K^-})_{pp}$&$(N_R^{K^-})_{pp}$&$(\beta_{K^-})_{pp}$\\
0.165&1.454&0.180\\
\hline
$(\alpha_{\bar p})_{pp}$&$(N_R^{\bar p})_{pp}$&$(\beta_{\bar p})_{pp}$\\
0.105&1.573&0.180\\
\hline
\hline
\end{tabular}
\end{center}
\end{table}
\begin{table}
\caption{Values of $(\alpha)_{PbPb}$, $(N_R)_{PbPb}$ and $(\beta)_{PbPb}$  for different centralities for the
$\pi^-$, $K^-$ and $\bar p$ productions in $Pb+Pb$ collisions at
$\sqrt{s_{NN}}$=2.76 TeV}
\begin{center}
\begin{tabular}{cccc}
\hline
\hline
Centrality&$(\alpha_{\pi^-})_{PbPb}$&($N_R^{\pi^-})_{PbPb}$&$(\beta_{\pi^-})_{PbPb}$\\
0-5\%& $5.8\times10^4$&2.493&0.30\\
5-10\%& $2.5\times10^4$&2.483&0.30\\
10-20\%&$8.5\times10^3$&2.472&0.30\\
20-30\%&$2.5\times10^3$&2.464&0.30\\
30-40\%&$6.8\times10^2$&2.456&0.30\\
40-50\%&$2.5\times10^2$&2.444&0.30\\
50-60\%&48&2.434&0.30\\
60-70\%&9.1&2.415&0.30\\
70-80\%&1.8&2.397&0.30\\
80-90\%&0.45&2.392&0.30\\
\hline
Centrality&$(\alpha_{K^-})_{PbPb}$&($N_R^{K^-})_{PbPb}$&$(\beta_{K^-})_{PbPb}$\\
0-5\%& $6.3\times10^3$&3.314&0.180\\
5-10\%& $3.1\times10^3$&3.015&0.180\\
10-20\%&$7.4\times10^2$&2.784&0.180\\
20-30\%&$1.8\times10^2$&2.583&0.180\\
30-40\%&46&2.321&0.180\\
40-50\%&12&2.124&0.180\\
50-60\%&3.2&2.111&0.180\\
60-70\%&0.62&2.104&0.180\\
70-80\%&0.14&2.008&0.180\\
80-90\%&0.034&1.984&0.180\\
\hline
Centrality&$(\alpha_{\bar p})_{PbPb}$&($N_R^{\bar p})_{PbPb}$&$(\beta_{\bar p})_{PbPb}$\\
0-5\%& $3.1\times10^3$&3.114&0.180\\
5-10\%& $8.4\times10^2$&2.723&0.180\\
10-20\%&$2.6\times10^2$&2.534&0.180\\
20-30\%&66&2.302&0.180\\
30-40\%&23&2.286&0.180\\
40-50\%&5.9&2.234&0.180\\
50-60\%&1.5&2.201&0.180\\
60-70\%&0.43&2.194&0.180\\
70-80\%&0.072&2.075&0.180\\
80-90\%&0.012&2.034&0.180\\
\hline
\hline
\end{tabular}
\end{center}
\end{table}
\newpage

\begin{figure}
\centering
\includegraphics[width=2.7in]{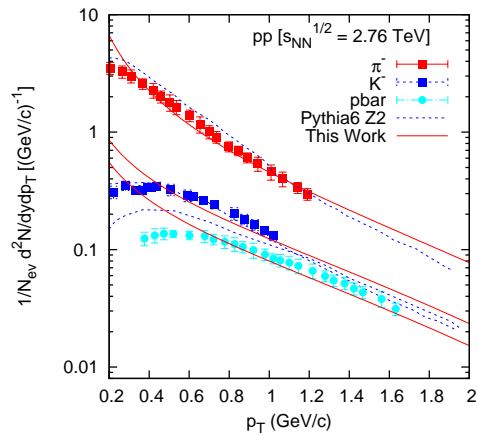}
  \caption{Plots for $\pi^-$, $K^-$, $\bar p$ and $p$ productions in $p+p$ collisions at
energies $\sqrt{s_{NN}}$ = 2.76 TeV. Data are taken \cite{cms12}. Solid lines in the Figures show the SCM-based
theoretical plots while the dotted ones show PYTHIA-based results \cite{cms12}.}
\end{figure}

\begin{figure}
\subfigure[]{
\begin{minipage}{.5\textwidth}
\centering
\includegraphics[width=2.5in]{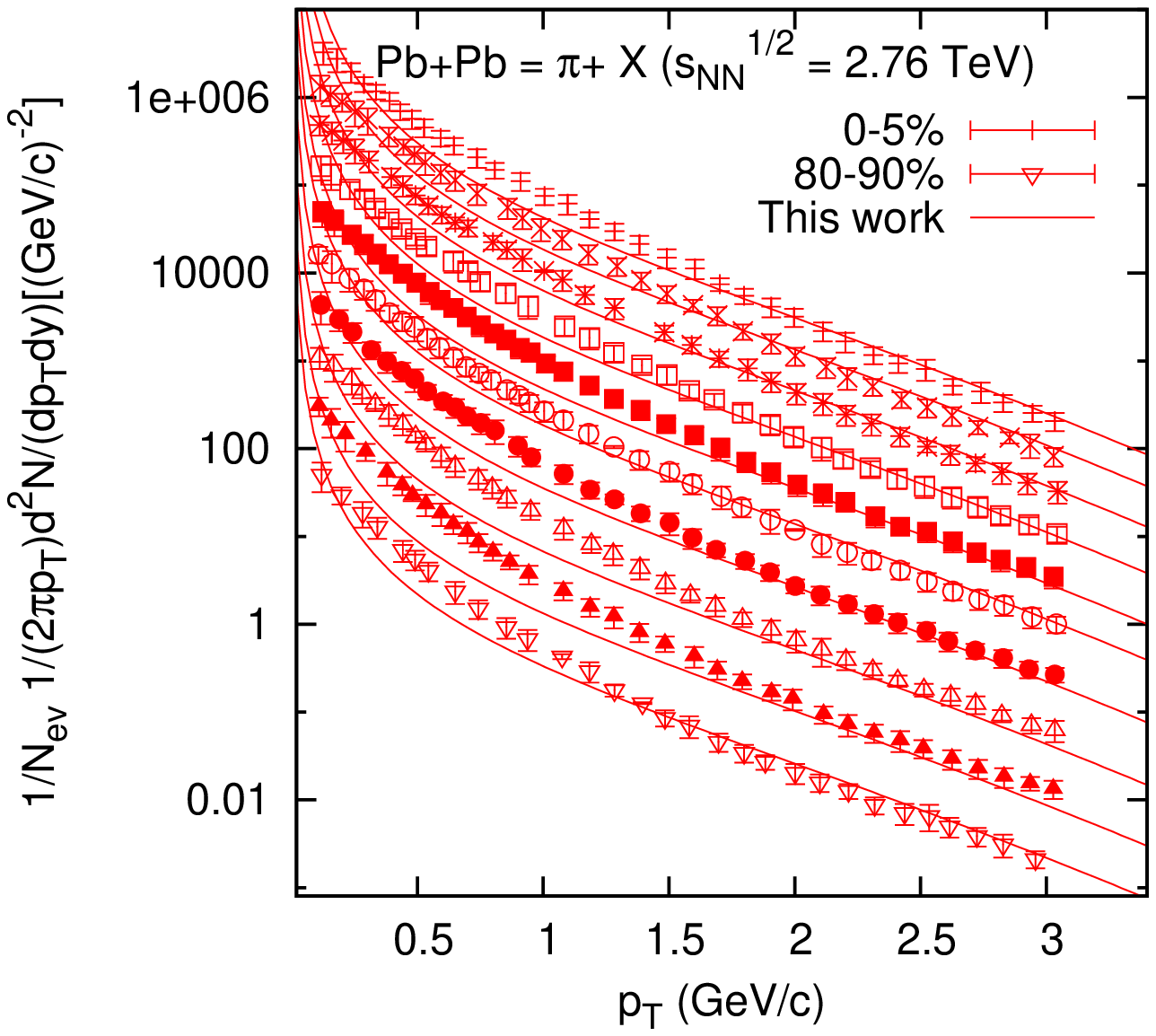}
\setcaptionwidth{2.3in}
\end{minipage}}%
\subfigure[]{
\begin{minipage}{0.5\textwidth}
\centering
 \includegraphics[width=2.5in]{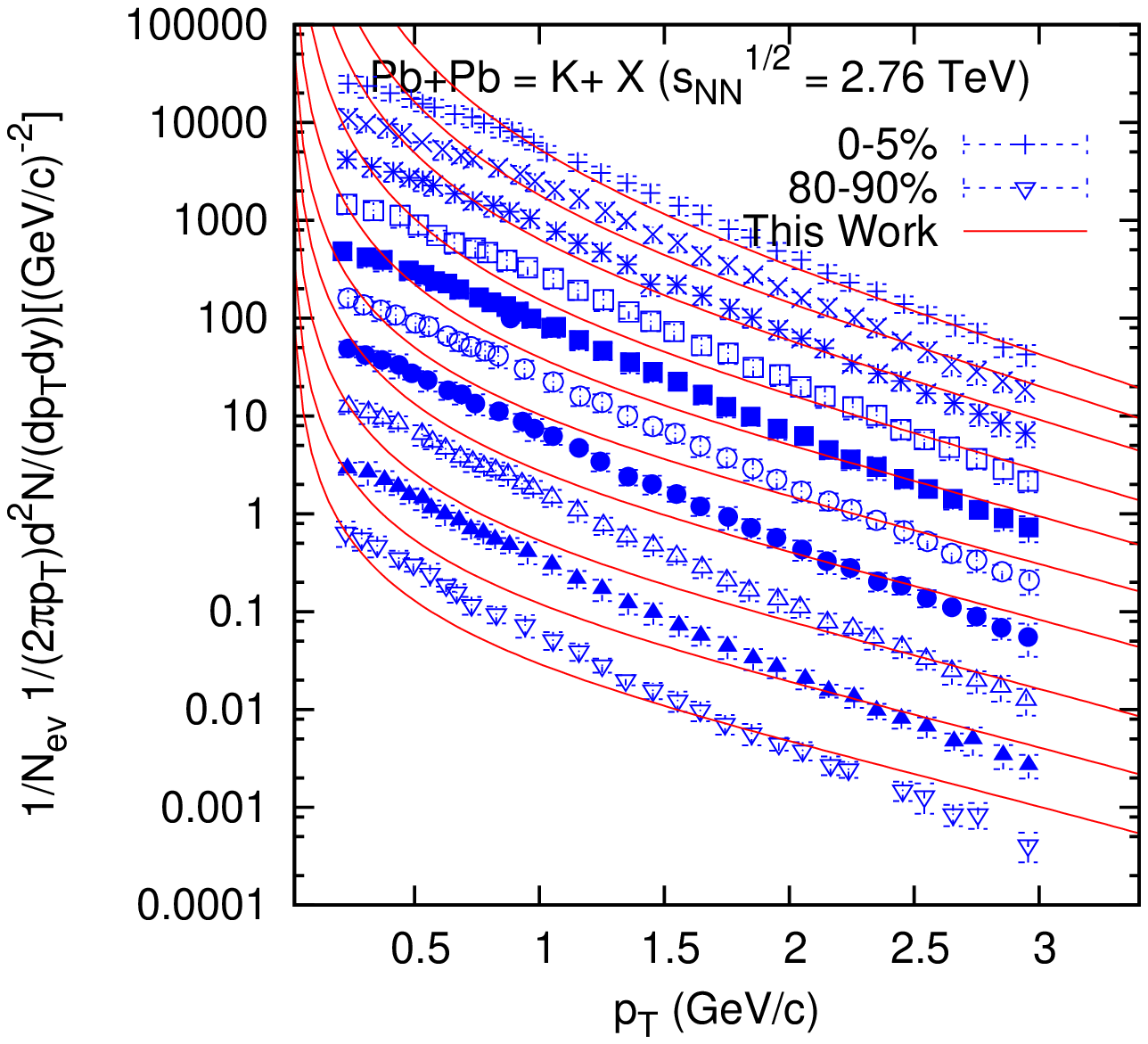}
  \end{minipage}}%
  \vspace{0.01in}
 \subfigure[]{
 \begin{minipage}{.5\textwidth}
\centering
 \includegraphics[width=2.5in]{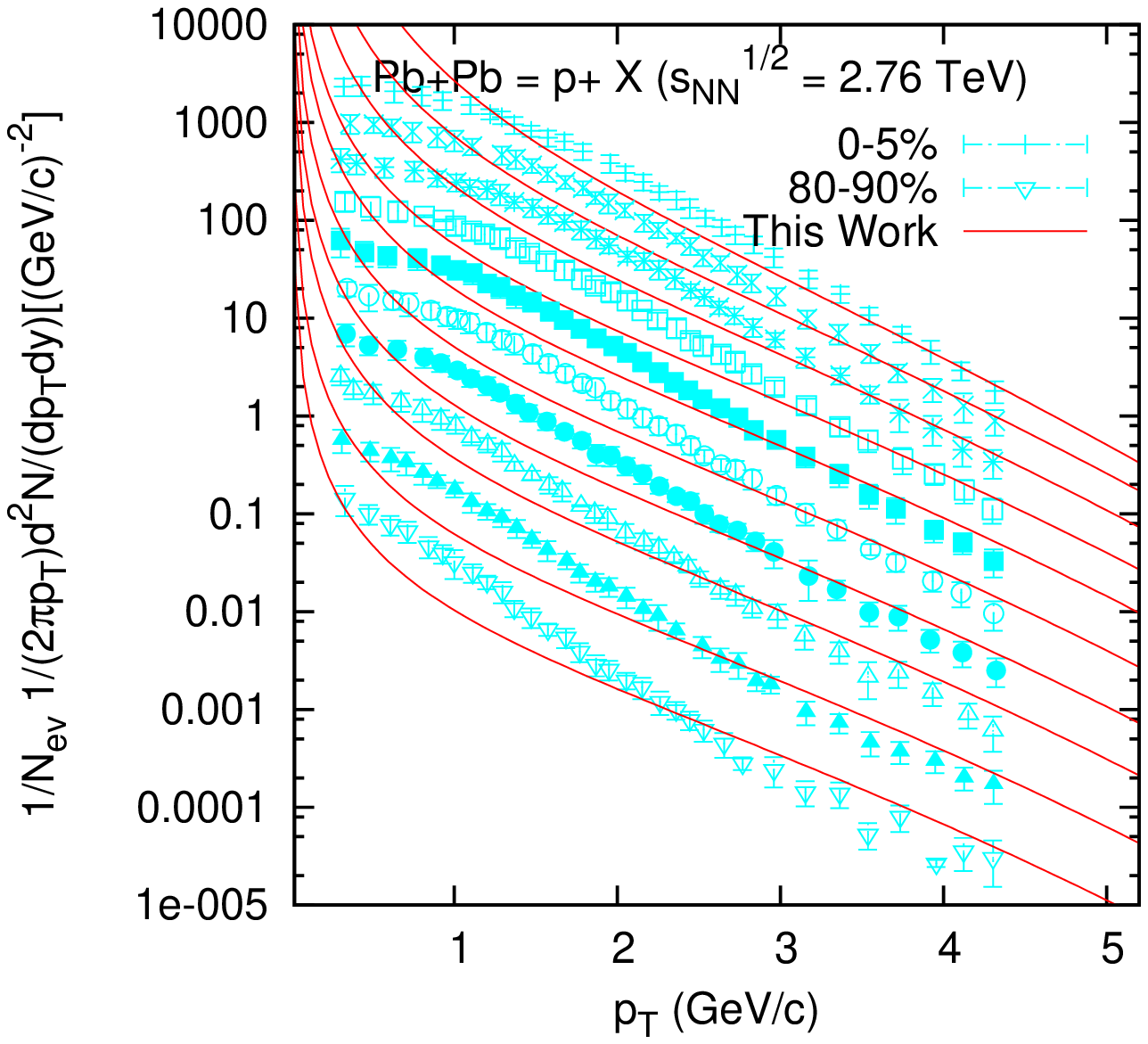}
 \setcaptionwidth{2.3in}
\end{minipage}}%
\caption{Centrality dependence of
the $p_T$ distribution for (a) $\pi^-$, (b) $K^-$ and (c) $\bar p$ for different centralities in $Pb+Pb$ collisions at
energies $\sqrt{s_{NN}}$ = 2.76 TeV. Data are taken from \cite{ALICE13}. The solid lines in
the Figures 2(a), 2(b) and 2(c) show the SCM
calculations for different centralities.}
\end{figure}

\begin{figure}
\subfigure[]{
\begin{minipage}{.5\textwidth}
\centering
\includegraphics[width=2.5in]{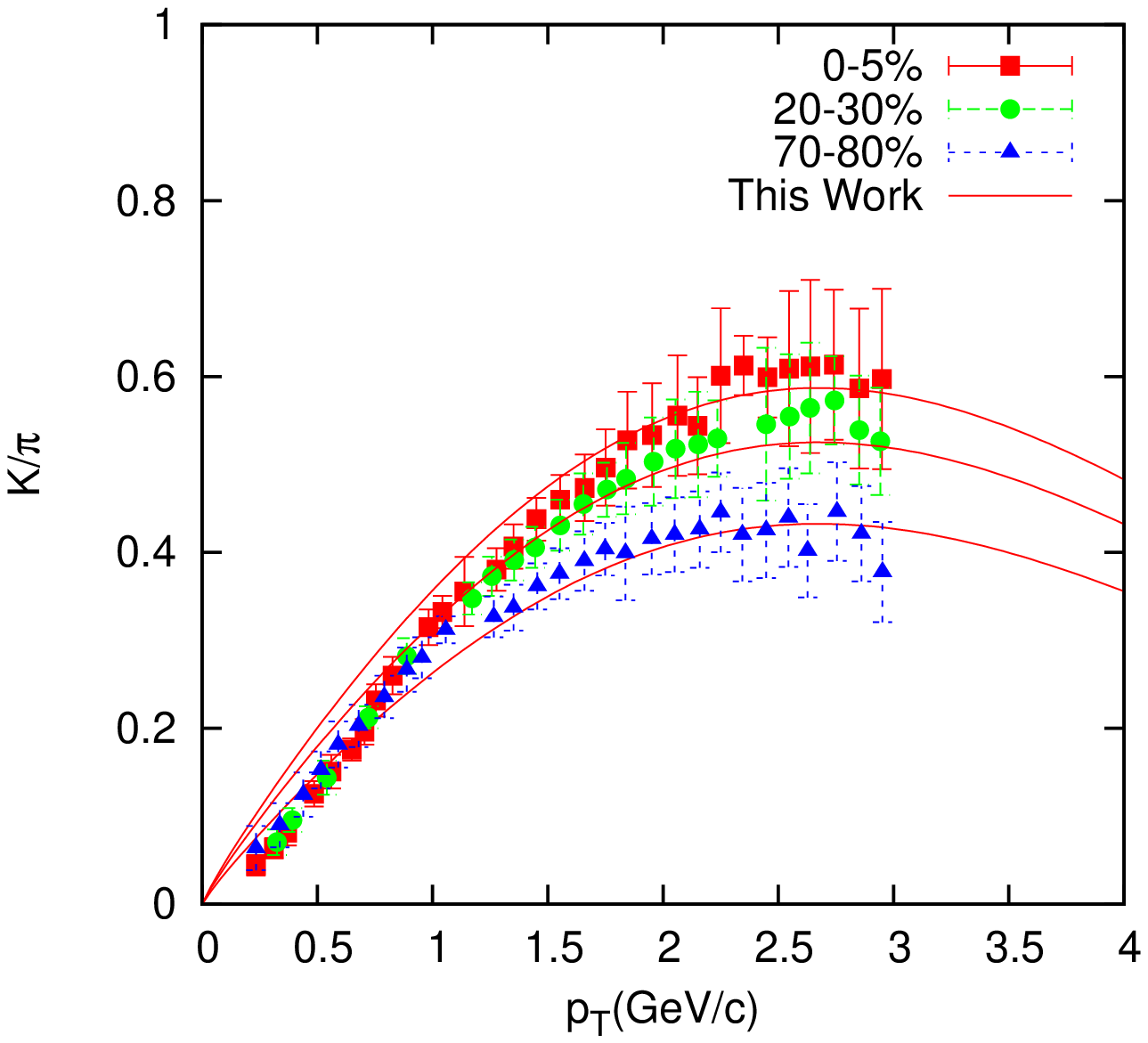}
\setcaptionwidth{2.3in}
\end{minipage}}%
\subfigure[]{
\begin{minipage}{0.5\textwidth}
\centering
 \includegraphics[width=2.5in]{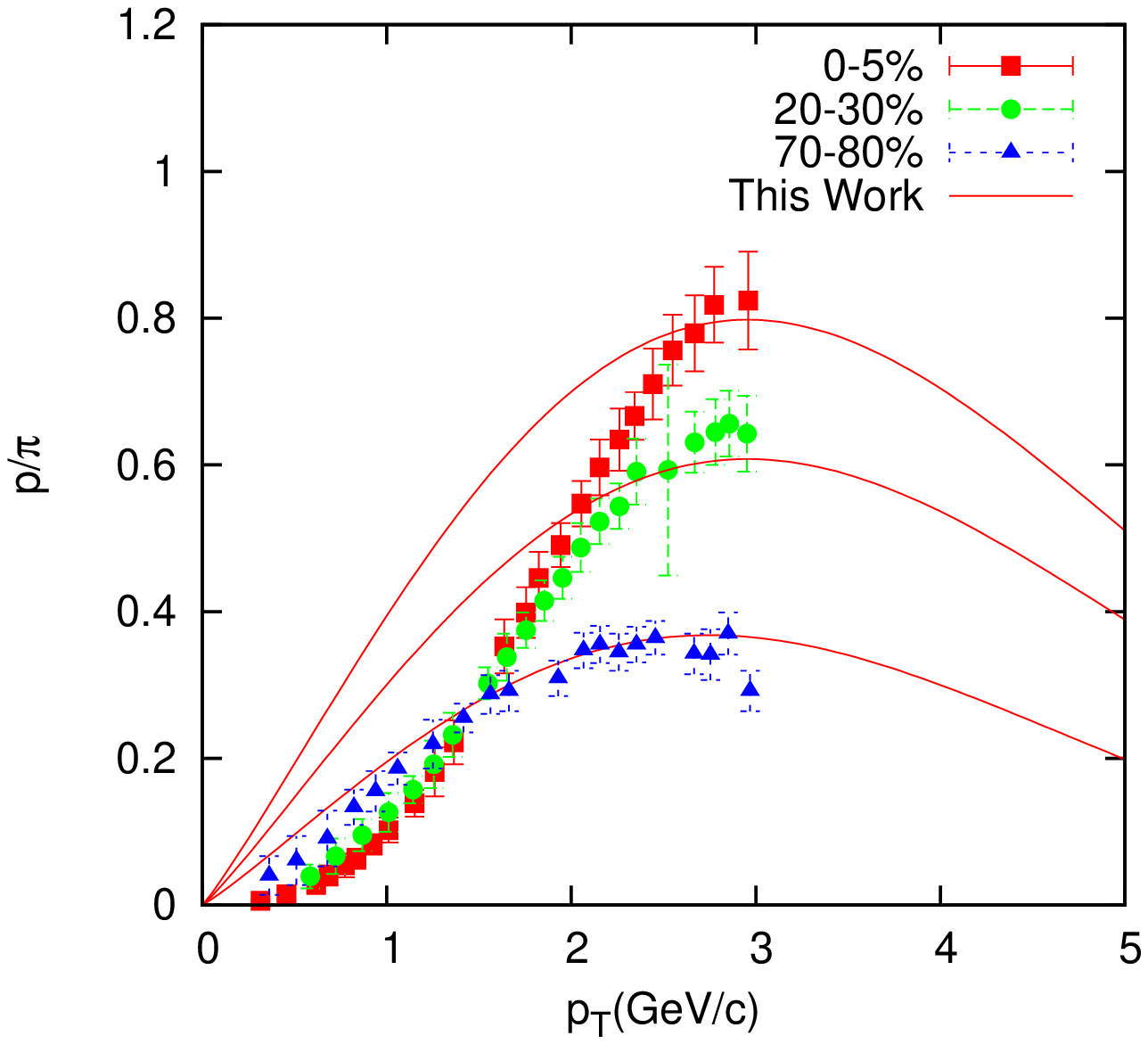}
  \end{minipage}}%
  \caption{Ratios of (a) $K/\pi$ and (b) $p/\pi$ as a
function $p_T$ for 0-5$\%$, 20-30$\%$ and 70-80$\%$ $Pb+Pb$ reactions at $\sqrt {s_{NN}}$
=2.76 TeV. Data in these Figures are taken from
\cite{ALICE13}. The solid lines show the SCM-based results.}
\end{figure}

\begin{figure}
\centering
\includegraphics[width=2.5in]{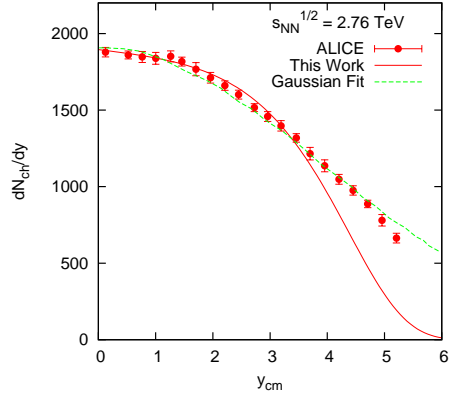}
   \caption{Plot of rapidity distribution of $\pi$ in central $Pb+Pb$ reactions at $\sqrt {s_{NN}}$
=2.76 TeV. Data in the Figure are taken from
\cite{ALICE132}. The solid line shows the SCM-based results while the dotted line depicts the Gaussian fit\cite{ALICE132}.  }
\end{figure}

\begin{figure}
\centering
\includegraphics[width=2.5in]{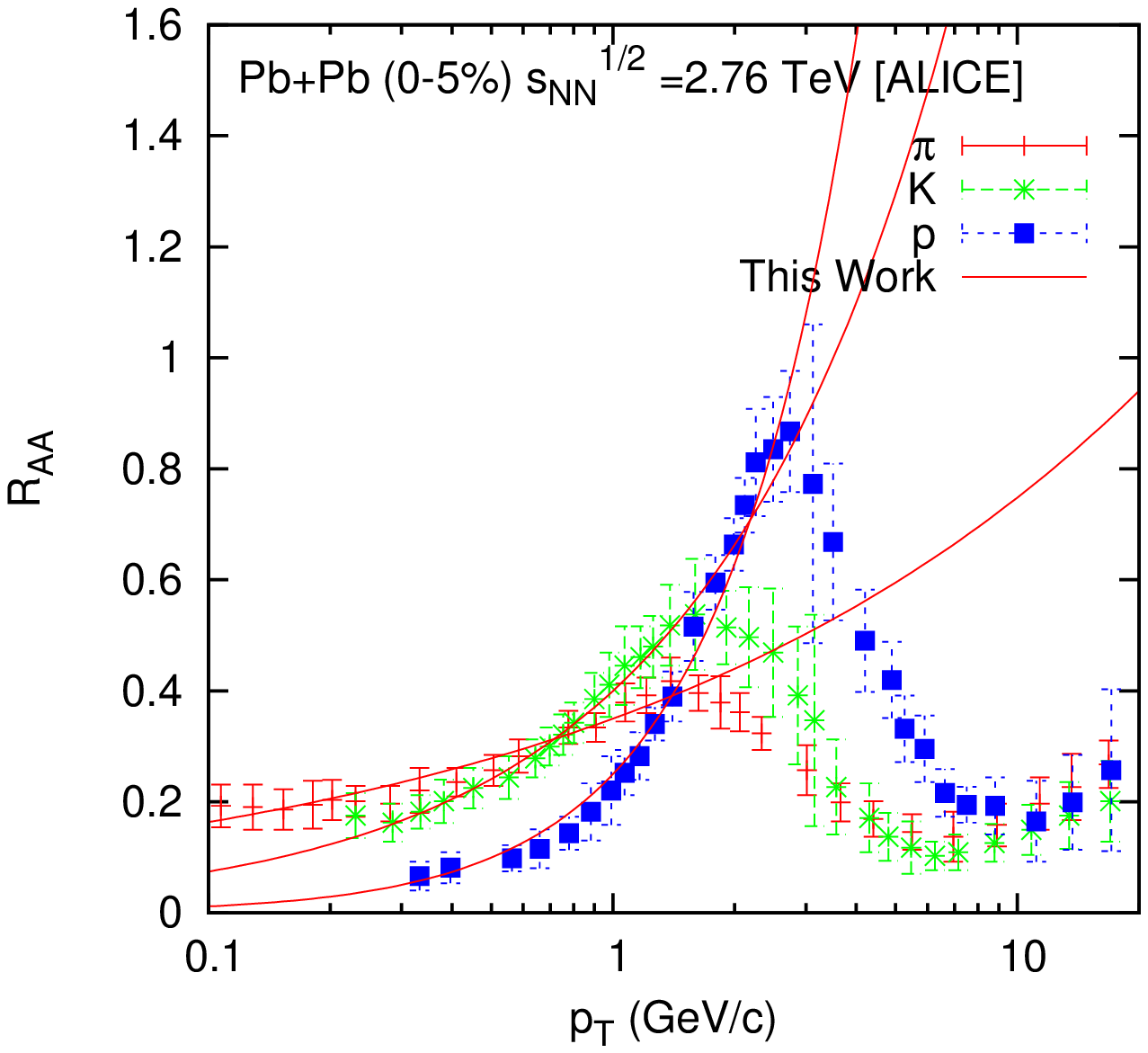}
  \caption{Plots for $R_{AA}$ versus $p_T$ in central $Pb+Pb$ collisions at
energies $\sqrt{s_{NN}}$ = 2.76 TeV. Data are taken from Ref. \cite{marek}. Solid lines in the Figure show the SCM-based
theoretical plots.}
\end{figure}

\newpage
\begin{thebibliography}{*}
\bibitem{alice1401} B. Abelev et al., [ALICE Collaboration], arXiv:1401.1250v2 [nucl-ex] (16 May 2014).
\bibitem{riggi} F. Riggi, J.  Phys. (Conference Series) {\bf{424}}, 012004 (2013).
\bibitem{zhang} S. Zhang , L. X. Han , Y. G. Ma , J. H.Chen and  C. Zhong, Phys. Rev C{\bf 89}, 034918 (2014).
\bibitem{deflo} D. de Florian et al., Phys. Rev. Letts. {\bf{113}}, 012001 (2014).
\bibitem{pgr14} P. Guptaroy, S. Guptaroy, Chin. Phys. Letts, {\bf{31}}, 082501, (2014); [arXiv:1406.6296v1 [hep-ph] 24 Jun 2014].
\bibitem{pgr12} P. Guptaroy, Goutam Sau, $\&$ S. Bhattacharyya,
 J. of Mod. Phys., {\bf{3}}, 116 (2012); [arXiv:1110.6612 v1 [hep-ph] 30 Oct 2011].
\bibitem{pgr10}  P. Guptaroy, G. Sau, S. K. Biswas, S. Bhattacharyya,  IL Nuovo Cimento B
{\bf{125}}, 1071 (2010); [arXiv:0907.2008 v2 [hep-ph] 4 Aug 2010].
\bibitem{pgr07}  P. Guptaroy, Bhaskar De, G. Sau, S. K. Biswas, S. Bhattacharyya, Int. J. Mod. Phys. A
{\bf{28}}, 5121 (2007).
\bibitem{bhat78} P. Bandyopadhyay and S. Bhattacharyya, IL Nuovo
Cimento A{\bf{43}}, 305 (1978).
\bibitem{bhat881} S. Bhattacharyya, IL Nuovo Cimento C{\bf{11}}, 51 (1988).
\bibitem{bhat882}  S. Bhattacharyya,  J. Phys. G{\bf{14}}, 9 (1988).
\bibitem{pgr08}  P. Guptaroy, G. Sau, S. K. Biswas, S. Bhattacharyya, Mod. Phys. Lett. A
{\bf 23}, 1031 (2008).
\bibitem{wong} C. Y. Wong:`Introduction to High-Energy Heavy Ion Collisions'
(World Scientific,1994).
\bibitem{na5002} M. C. Abreu  et al., NA50 Collaboration Preprint, 15 Feb., 2002 CERN-EP/2002-017
\bibitem{cms12} S. Chatrchyan et al., [CMS Collab.] Eur. Phys. J. C{\bf{72}}, 2164 (2012).
\bibitem{ALICE13} B. Abelev et al., [ALICE Collab.] Phys. Rev. C{\bf{88}}, 044910 (2013).
\bibitem{abk} A. B. Kaidalov, K. A. Ter-Martirosyan, Sov. J. Nucl. Phys. {\bf{36}}, 979 (1984).
\bibitem{ALICE132} B. Abelev et al., [ALICE Collaboration], Phys. Lett. B {\bf{726}}, 610 (2013).[arXiv:1304.0347v2 [nucl-ex] 27 Jul 2013].
\bibitem{jacek} J. Otwinowski, PoS ConfinementX, 170 (2012); [arXiv:1301.5285v1 [hep-ex] (22 Jan 2013)].
\bibitem{marek} M. Chojnacki for the ALICE Collaboration, J. Phys.: Conf. Series {\bf{509}}, 012041 (2014).
\end{thebibliography}
\end{document}